# Giant Unruh effect in hyperbolic metamaterial waveguides

## Igor I. Smolyaninov

*Department of Electrical and Computer Engineering, University of Maryland, College Park, MD 20742, USA*
*smoly@umd.edu*

**The Unruh effect is the prediction that an accelerating object perceives its surroundings as a bath of thermal radiation even if it accelerates in vacuum. The Unruh effect is believed to be very difficult to observe in the experiment, since an observer accelerating at g=9.8 m/s² should see vacuum temperature of only 4x10⁻²⁰ K. Here we demonstrate that photons in metamaterial waveguides may behave as massive quasi-particles accelerating at up to 10²⁴ g, which is about twelve orders of magnitude larger than the surface acceleration near a stellar black hole. These record high accelerations may enable experimental studies of the Unruh effect and the loss of quantum entanglement in strongly accelerated reference frames.**

It is well established in the scientific literature (see for example [1-3]) that a photon in a waveguide behaves as a massive quasi-particle which may be characterized by both inertial and gravitational mass obeying the Einstein equivalence principle. For example, as illustrated in Fig. 1, photon propagation in a vertical waveguide subjected to Earth gravitational field is fully analogous to a vertical motion of a heavy body [2]. Indeed, let us consider an empty rectangular optical waveguide shown in Fig. 1(b) and assume that this waveguide has constant dimensions *d* and *b* in the *x*- and *y*- directions, respectively. Let us also assume that all walls of the waveguide are made of an ideal metal. The dispersion law of photons propagating inside this waveguide looks like a dispersion law of a massive quasi-particle:

$$\frac{\omega^2}{c^2} = k_z^2 + \frac{\pi^2 I^2}{d^2} + \frac{\pi^2 J^2}{b^2} \quad (1)$$

where $\omega$ is the photon frequency, $k_z$ is the photon wave vector in the *z*-direction, and *I* and *J* are the mode numbers in the *x*- and *y*- directions, respectively. The effective inertial rest mass of the photon in the waveguide is

$$m_{eff} = \frac{\hbar \omega_{ij}}{c^2} = \frac{\hbar}{c}\left(\frac{\pi^2 I^2}{d^2} + \frac{\pi^2 J^2}{b^2}\right)^{1/2} \quad (2)$$

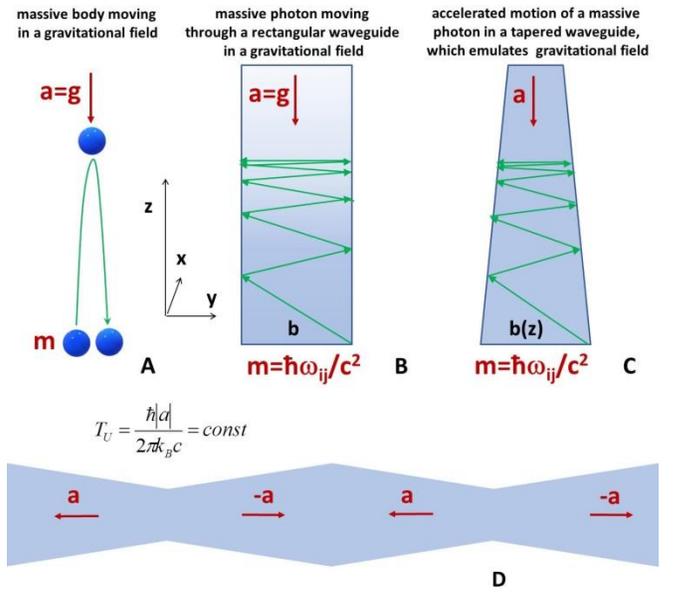

Fig. 1. Photon in a waveguide behaves as a massive quasi-particle: (a) A massive body thrown up in a gravitational field eventually reaches zero velocity and turns around. (b) Similar to a massive body, photon in a waveguide moving against an external gravitational field eventually stops and turns around near the waveguide cut-off. Gradient of the effective refractive index n in the waveguide is illustrated by shading (n=1 at the bottom of the waveguide) (c) The effect of external gravitational field on a photon in a waveguide may be emulated by waveguide tapering, which also leads to accelerated motion of a massive photon. (d) Periodically tapered waveguide geometry may be used to enable photon mode thermalization while keeping the Unruh temperature constant along the waveguide.

which is equal to its effective gravitational mass [2]. Let us assume that this waveguide is either immersed in a weak constant vertical gravitational field, or is subjected to accelerated motion with constant acceleration (these are locally equivalent situations). Since the gravitational field is static, this geometry may be represented by a static waveguide filled with a medium, in which the electric permittivity $\varepsilon$ and the magnetic permeability $\mu$ gradually change as a function of z-coordinate. Indeed, the equations of electrodynamics in the presence of static gravitational field look exactly like Maxwell equations in the medium in which $\varepsilon = \mu = g_{00}^{-1/2}$, where $g_{00}$ is the $tt$ component of the metric tensor [4]. Since the gravitational field is weak (there is no event horizon),

$$g_{00} \approx 1 + \frac{2\phi}{c^2} \quad (3)$$

where the gravitational potential $\phi = gz$ [4]. Thus, the waveguide subjected to a gravitational field (or experiencing a constant acceleration) may be represented by a waveguide filled with a medium in which $\varepsilon = \mu$ and both $\varepsilon$ and $\mu$ have a gradient in the $z$-direction, so that

$$n = (\varepsilon\mu)^{1/2} \approx 1 - \frac{gz}{c^2} \quad (4)$$

In other words, the "optical dimensions" of the waveguide ($nd$ and $nb$) change as a function of the $z$-coordinate. If the conventional optical terminology is used, such a waveguide is called a tapered waveguide. As illustrated in Fig. 1, similar to a massive body, photon in this waveguide moving against an external gravitational field eventually stops and turns around near the waveguide cut-off. Note that the effective refractive index gradient may be related to the waveguide acceleration as

$$a = g = -c^2 \frac{dn}{dz} \quad (5)$$

This expression is fully analogous to the definition of an acceleration of a heavy body $a = dv/dt = vdv/dz$, if $v$ is identified as the group velocity of the photon in the waveguide $c_{gr} = d\omega/dk_z = nc$.

Let us now consider an adiabatically tapered rectangular waveguide shown in Fig. 1(c), which $b$ dimension changes slowly as a function of $z$. For the sake of simplicity let us initially assume that the waveguide is empty, the walls of the waveguide are made of an ideal metal, the width $d$ of the waveguide in the $x$-direction is kept constant, and $d \ll b$. If a photon is coupled into the $l=1$ mode of this adiabatically tapered waveguide, it is going to stay in this mode, and its effective mass

$$m_{eff} \approx \frac{\hbar\pi}{cd} \quad (6)$$

remains constant. Moreover, the slow changes in $b(z)$ will lead to the photon staying in the same transverse mode $J$ of the waveguide during its adiabatic propagation [3]. Therefore, the dispersion law of such a photon may be written as

$$\frac{\omega^2}{c^2} = k_z^2 + \frac{\pi^2}{d^2} + \frac{\pi^2 J^2}{b^2(z)}, \quad (7)$$

and its acceleration may be calculated [3] as

$$a = c_{gr}\frac{dc_{gr}}{dz} = \frac{\pi^2 J^2 c^4}{\omega^2 b^3}\left(\frac{db}{dz}\right) \approx c^2 \frac{J^2 \lambda_0^2}{4b^3}\left(\frac{db}{dz}\right), \quad (8)$$

where $\lambda_0 = 2\pi\chi/\omega$ is the photon wavelength in free space.

Recently it was pointed out [3] that the photon acceleration in a conventional dielectric waveguide may reach very large magnitudes. Indeed, let us use Eq.(5) to evaluate typical photon accelerations achievable in a gradient index waveguide. In order for the quasi-classical ray optics approximation to remain valid, we need to assume that the refractive index variations remain slow ($\Delta n \leq 0.1n$) across an element of waveguide length $\Delta z$, which is much larger than the photon wavelength $\lambda_0/n$ inside the waveguide. These assumptions result in the following estimate of the achievable photon acceleration:

$$a \propto 10^{-2} c^2 \frac{n^2}{\lambda_0} \propto 10^{22} m/s^2 \propto 10^{21} g, \quad (9)$$

where $\Delta n \sim 0.1n$, $n \sim 2$, and $\Delta z \sim 10\lambda_0/n$ has been substituted into Eq.(5), and $\lambda_0 \sim 400$ nm has been assumed as the free space light wavelength. According to Eq.(8), a similar acceleration magnitude $\sim 0.1c^2/\lambda_0$ may also be obtained in a conventional adiabatically tapered waveguide, assuming slow $db/dz \sim 0.1$ variations, and the transverse waveguide dimensions $b(z)$ of the order of the free space light wavelength $\lambda_0$.

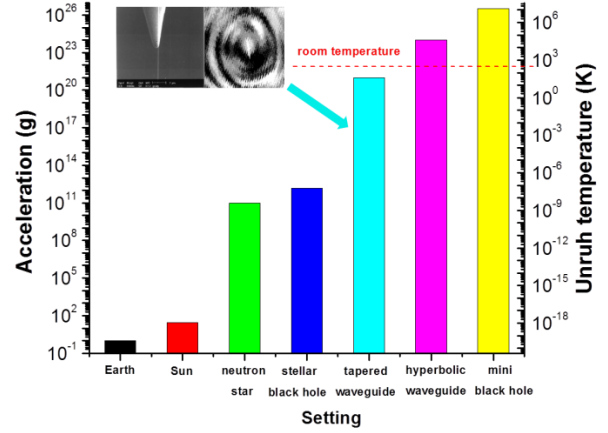

Fig. 2. Comparison of accelerations achievable in different astronomical and terrestrial settings. The corresponding Unruh temperatures are shown on the right side of the plot. The inset illustrates observations of mode coupling in a conventional tapered waveguide.

However, the most striking effects of such a strongly accelerated motion, such as the Unruh effect, which predicts that an accelerating object perceives its surroundings as a bath of thermal radiation even if it accelerates in vacuum [5], and the loss of quantum entanglement due to strongly accelerated motion [6], remained mostly out of reach even at such high accelerations. Indeed, the Unruh effect is believed to be very difficult to observe in the experiment, since an observer accelerating at $g=9.8$ m/s$^2$ should see vacuum temperature of only $4\times10^{-20}$ K. As illustrated in Fig. 2, even at $10^{21}g$ the Unruh temperature

$$T_U = \frac{\hbar a}{2\pi k_B c} \quad (10)$$

remains considerably below the room temperature, which makes observations very difficult. As far as the quantum entanglement loss due to accelerated motion is concerned, there was a recent attempt to verify this prediction in the experiment [7]. However, the result was negative due to very limited (<30$g$) range of experimentally accessible accelerations.

In this paper we will demonstrate that much larger accelerations of up to $10^{24}g$ may be obtained in metamaterial waveguides, which may enable not only the Unruh effect observation, but may also be used to study the physics of record high gravitational fields in terrestrial labs. For example, the proposed waveguide geometries appear to be a natural choice for the experiments on quantum entanglement loss due to accelerated motion. Compared to the recent experiments described in [7], the dynamic range of available accelerations would be increased by twenty two orders of magnitude. As illustrated in Fig. 2, these kinds of accelerations are about twelve orders of magnitude larger than the surface acceleration near a stellar black hole. Similar kinds of accelerations may only exist near the hypothesized microscopic black holes [8].

Analysis of Eqs. (5) and (8) suggests that in order to achieve even higher photon accelerations compared to the conventional optical waveguides depicted in Fig. 1, we need to decrease considerably both the waveguide dimensions and the photon wavelength inside the waveguide, so that the quasi-classical ray optics approximation may remain valid. This may be achieved in a waveguide made of recently developed hyperbolic metamaterials [9-13]. Hyperbolic metamaterials are extremely anisotropic uniaxial electromagnetic materials, which behave like a metal in one direction and like a dielectric in the orthogonal direction. Recent reviews of their properties may be found in [14,15]. Let us consider an adiabatically tapered rectangular waveguide, which walls are made of an ideal metal, and which is filled with a hyperbolic metamaterial having $\varepsilon_y = \varepsilon_1 > 0$ and $\varepsilon_x = \varepsilon_z = \varepsilon_2 < 0$. The dispersion law of extraordinary photons propagating inside this waveguide looks similar to Eq.(1):

$$\frac{\omega^2}{c^2} = \frac{k_z^2}{\varepsilon_1} + \frac{\pi^2 I^2}{\varepsilon_1 d^2} + \frac{\pi^2 J^2}{\varepsilon_2 b^2}, \qquad (11)$$

However, the different signs of $\varepsilon_1$ and $\varepsilon_2$ strongly affect the available range of the photon wavelengths inside the waveguide. Unlike in conventional dielectric materials, there is no diffraction limit on $k_z$ in the hyperbolic metamaterial waveguide, so that its $d$ and $b$ dimensions (along the $x$ and $y$-direction, respectively) may be made much smaller than the free space light wavelength $\lambda_0$. Note also that the positive sign of the effective mass of the photon in such a waveguide may still be maintained by the proper choice of $I$ and $J$ mode numbers.

For the sake of simplicity let us once again assume that $d$ is kept constant, while the $b$ dimension of the waveguide changes slowly as a function of $z$. The photon acceleration in such a waveguide may be calculated similar to Eq. (8) as

$$a = c_{gr} \frac{dc_{gr}}{dz} = \frac{\pi^2 J^2 c^4}{\varepsilon_1 \varepsilon_2 \omega^2 b^3} \left(\frac{db}{dz}\right) = c^2 \frac{J^2 \lambda_0^2}{4\varepsilon_1 \varepsilon_2 b^3} \left(\frac{db}{dz}\right) \qquad (12)$$

Since $b<<\lambda_0$ conditions are possible to achieve in a hyperbolic metamaterial waveguide, Eq. (12) indicates that much larger photon accelerations may be obtained in a tapered hyperbolic waveguide compared to the conventional one. Let us evaluate the realistic upper bound on these accelerations.

In the absence of the conventional diffraction limit [10], this bound is set by the size of the hyperbolic metamaterial unit cell $s$, introducing a natural wave number cut-off:

$$k_{max} \sim 1/s \qquad (13)$$

Depending on the metamaterial design and the fabrication method used, the unit cell size in artificial optical hyperbolic metamaterials runs from $s\sim10$ nm in semiconductor-dielectric [16] and metal-dielectric [13] layered structures, to $s\sim100$ nm in nanowire composites [17,18]. Very recently it was also pointed out that quite a few natural materials also exhibit hyperbolic properties in the visible range [19]. This sets $s\sim1$ nm limit on the hyperbolic material unit cell. Choosing a more conservative and realistic estimate $b\sim70$ nm, which coincides with the observed spatial resolution of a hyperlens [12,13], leads to the upper bound of $a\sim10^{24}g$ on the achievable photon acceleration in a hyperbolic metamaterial waveguide (again assuming $db/dz\sim0.1$). As illustrated in Fig. 2, the corresponding Unruh temperature perceived by a photon accelerating in a hyperbolic metamaterial waveguide reaches up to $\sim30000$ K, which is considerably higher than the room temperature, and has about the same order of magnitude as the photon energy $\hbar\omega$ in the visible range. Let us discuss how these giant accelerations may be used to study such non-trivial effects as loss of quantum entanglement and the Unruh effect in accelerated reference frames.

As was noted in [3], mode coupling in a waveguide may be used as a "thermometer" to measure the Unruh temperature. An example of such measurements in a conventional tapered waveguide is shown in the inset in Fig. 2. Initially, laser light may be coupled only into some particular optical mode $(k_{z0}, I_0, J_0)$ of a multimode waveguide. Due to light scattering by the volume and the surface defects inside the waveguide, massive photon quasi-particles gradually populate the other guided modes available for the photons at the same light frequency $\omega$. Thus, in the reference frame co-moving with the photon quasi-particles, two kinds of effective temperatures may be introduced for this system: the temperature $T_k$ of the lateral 1D motion of the massive photon quasi-particles along the $z$-direction, and the temperature $T_{ij}$ of the internal $(I,J)$ degrees of freedom of the photons, which corresponds to the mode coupling in the waveguide, and which is similar to the spin temperature of a nuclear spin system [20]. When the laser light is initially coupled into some particular waveguide mode, both temperatures are zero. However, upon propagation both $T_k$ and $T_{ij}$ must equalize with the waveguide temperature. It was proposed in [3] that the mode coupling in a conventional tapered waveguide may potentially be used to observe the Unruh effect. It is important to note that such an observation scheme does not require any additional "energy source" inside the tapered waveguide since the total energy of the photon system in this scheme (and the energy of each photon) remains constant in the laboratory reference frame. However, since the achievable Unruh temperatures in such waveguides were considerably below the room temperature (see Fig. 2), experimental measurements appeared to be extremely difficult [3].

Considerable enhancement of the Unruh temperature in the tapered hyperbolic metamaterial waveguides presents us with an opportunity to do such an experiment. Indeed, massive photon quasi-particles, which experience accelerated motion through a tapered hyperbolic waveguide, will perceive the waveguide temperature as being equal to the Unruh temperature. As a result, upon propagation both $T_k$ and $T_{ij}$ temperatures are supposed to equalize at the Unruh temperature, which will be much higher than the room temperature in the laboratory reference frame. Since such a thermalization may require an extended propagation time, and it may not be achieved inside a single taper, periodically tapered waveguide geometry may be used as illustrated in Fig. 1(d). This is possible because the Unruh temperature depends only on the magnitude of acceleration, and does not depend on its direction (see Eq. (10)). Evaluation of the degree of mode coupling at the output face of the waveguide will enable measurements of the Unruh temperature experienced by the photons. Such an experiment would supplement recent observation of classical analog of Unruh effect in a water-wave system [21]. We should also mention that recent studies [22] demonstrated that the issue of propagation losses in hyperbolic metamaterials may be managed using gain media as a dielectric component of the metamaterial, so that the hyperbolic metamaterial waveguide may be made almost lossless in the narrow spectral region around the photon frequency $\omega$ considered above. Other experimental difficulties related to the waveguide defects and asymmetries (as described in detail in [3]) must also be carefully addressed.

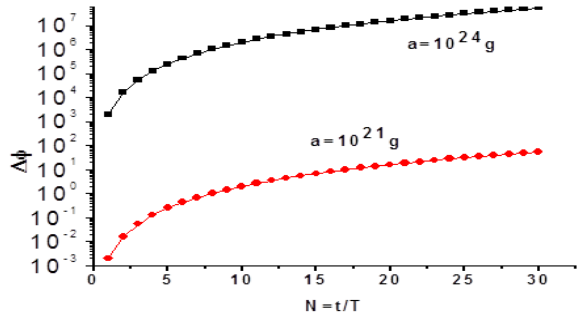

Fig. 3. The relative phase factor (Eq.(15)) plotted for $\nu=8\times10^{14}$ Hz as a function of propagation time expressed as a number of light oscillation periods $N=t/T$ for the cases of conventional tapered waveguide ($a\sim10^{21}g$) and hyperbolic metamaterial waveguide ($a\sim10^{24}g$).

Another consequence of the Unruh effect is that an entangled pure state seen by inertial observers appears mixed from an accelerated reference frame. As a result, a two qubit state which is maximally entangled in an inertial frame becomes less entangled if the observers are relatively accelerated. In the limit of infinite acceleration, which can be applied to the situation of one of the observers falling into a black hole while the other barely escapes, the distillable entanglement vanishes [6]. Quantum entangled accelerating three qubit states also show entanglement loss [23]. An attempt to verify these predictions in the experiment has been reported very recently [7]. However, these experiments produced negative result due to very limited (<30g) range of experimentally accessible accelerations. The proposed tapered waveguide geometries appear to be a natural choice for more sensitive experiments on quantum entanglement loss due to accelerated motion. Compared to the experiments described in [7], the dynamic range of available accelerations would be increased by about twenty two orders of magnitude.

Recent theoretical results [24] indicate a clear quantitative and operational connection between quantum coherence and entanglement. Therefore, let us consider the effect of accelerated motion of photons in a tapered waveguide on the loss of quantum coherence. Based on the equivalence principle, we could pursue two alternative ways to take into account an external gravitational field (or accelerated motion) on the photons in the waveguide. The first, the Newtonian perspective, would simply be to incorporate a term in the Hamiltonian representing the Newtonian potential, and use standard Cartesian coordinates (x,y,z,t). The second, the Einsteinian perspective, would be to adopt a freely falling reference system (X,Y,Z,T), in accordance with which the external gravitational field vanishes. Let us denote the photon wavefunction in the (x,y,z,t) system, using the Newtonian perspective, by $\psi$, whereas for the (X,Y,Z,T) system, using the Einsteinian perspective, we will use $\Psi$. To get consistency between the two perspectives, we need to relate $\psi$ to $\Psi$ by a phase factor [25]:

$$\Psi = e^{i\frac{m}{\hbar}\left(\frac{1}{6}t^3 a^2 - taz\right)}\psi \qquad (14)$$

It is important to note that this phase factor is quite meaningful, since it contains a nonlinear time-dependence $\sim t^3 a^2$, and therefore it affects the splitting of field amplitudes into positive and negative frequencies. In other words, the Einsteinian and Newtonian wavefunctions belong to different Hilbert spaces, corresponding to different quantum field theoretic vacua. In fact, this situation is just another manifestation of the Unruh effect, which was described earlier (see Eq.(10)). It is also easy to see that this phase factor leads to very fast loss of quantum coherence at large propagation times $t$ and at large accelerations. Indeed, using Eq.(2) the phase factor in Eq.(14) may be written as

$$\Delta\phi = \frac{\omega_{ij} t^3 a^2}{6c^2} \qquad (15)$$

This phase factor is plotted in Fig.3 as a function of propagation time expressed as a number of light oscillation periods $N=t/T$ for the cases of conventional tapered waveguide ($a\sim10^{21}g$) and hyperbolic metamaterial waveguide ($a\sim10^{24}g$). The loss of coherence ($\Delta\phi\sim1$) appears to occur very fast in both cases.

Thus, the record high accelerations, which may be created using hyperbolic metamaterial waveguides in terrestrial laboratories, may enable experimental studies of loss of quantum entanglement and the Unruh effect in accelerated reference frames. Due to the Einstein equivalence principle, similar settings may also be used to study the physics of record high gravitational fields. Some potential complications may however arise from the fact that local acceleration is equivalent to a local gravitational field, but this is not true globally, as discussed in [26].